\documentclass[preprint2]{aastex63} 

\newcommand{\epsind}{\texorpdfstring{$\varepsilon$}{epsilon}~Indi~A}
\newcommand{\numax}{\mbox{$\nu_{\rm max}$}}
\newcommand{\dnu}{\mbox{$\Delta \nu$}}
\newcommand{\Dnu}{\dnu}
\newcommand{\muhz}{\mbox{$\mu$Hz}}
\newcommand{\muHz}{\muhz}

\usepackage[amssymb]{SIunits} 

\usepackage{CJKutf8}
\newcommand{\CNnames}[1]{{\begin{CJK}{UTF8}{gbsn}~(#1)~\end{CJK}}}

\begin{document}

\title{Low-amplitude solar-like oscillations in the K5\,V star $\varepsilon$ Indi A}

\correspondingauthor{Mia S. Lundkvist}
\email{lundkvist@phys.au.dk}

\author[0000-0002-8661-2571]{Mia S. Lundkvist} 
\affiliation{Stellar Astrophysics Centre, Department of Physics and Astronomy, Aarhus University, 8000 Aarhus C, Denmark}

\author[0000-0002-9037-0018]{Hans Kjeldsen} 
\affiliation{Stellar Astrophysics Centre, Department of Physics and Astronomy, Aarhus University, 8000 Aarhus C, Denmark}
\affiliation{Aarhus Space Centre (SpaCe), Department of Physics and Astronomy, Aarhus University, Denmark}

\author[0000-0001-5222-4661]{Timothy R. Bedding} 
\affiliation{Sydney Institute for Astronomy, School of Physics, University of Sydney NSW 2006, Australia}

\author[0000-0002-1452-5268]{Mark J. McCaughrean} 
\affiliation{European Space Agency, ESTEC, Postbus 299, 2200 AG Noordwijk, The Netherlands}

\author[0000-0003-1305-3761]{R. Paul Butler} 
\affiliation{Earth and Planets Laboratory, Carnegie Institution for Science, 5241 Broad Branch Road, NW, Washington, DC 20015-1305, USA}

\author[0000-0003-4538-9518]{Ditte Slumstrup} 
\affiliation{European Southern Observatory, Alonso de Cordova 3107, Vitacura, Santiago, Chile}
\affiliation{Instituto de Estudios Astrof\'isicos, Facultad de Ingenier\'ia y Ciencias, Universidad Diego Portales, Av. Ej\'ercito Libertador 441, Santiago, Chile}

\author[0000-0002-4588-5389]{Tiago L. Campante} 
\affiliation{Instituto de Astrof\'{\i}sica e Ci\^{e}ncias do Espa\c{c}o, Universidade do Porto,  Rua das Estrelas, 4150-762 Porto, Portugal}
\affiliation{Departamento de F\'{\i}sica e Astronomia, Faculdade de Ci\^{e}ncias da Universidade do Porto, Rua do Campo Alegre, s/n, 4169-007 Porto, Portugal}

%
%
\author[0000-0003-1822-7126]{Conny Aerts} 
\affiliation{Institute of Astronomy, Department of Physics and Astronomy, KU Leuven, Celestijnenlaan 200 D, 3001 Leuven, Belgium}

\author[0000-0002-4696-6041]{Torben Arentoft} 
\affiliation{Stellar Astrophysics Centre, Department of Physics and Astronomy, Aarhus University, 8000 Aarhus C, Denmark}

\author{Hans Bruntt} 
\affiliation{Stellar Astrophysics Centre, Department of Physics and Astronomy, Aarhus University, 8000 Aarhus C, Denmark}

\author{C\'atia V. Cardoso} 
\affiliation{Aurora Technology B.V. for ESA, ESTEC, Keplerlaan 1, 2200 AG Noordwijk, The Netherlands}

\author[0009-0005-1167-0311]{Fabien Carrier} 
\affiliation{Institute of Astronomy, Department of Physics and Astronomy, KU Leuven, Celestijnenlaan 200 D, 3001 Leuven, Belgium}

\author[0000-0002-2167-8246]{Laird M. Close} 
\affiliation{Department of Astronomy, University of Arizona, 933 N. Cherry Ave., Tucson, AZ 85718, USA}

\author[0000-0001-8056-9202]{Jo{\~a}o Gomes da Silva} 
\affiliation{Instituto de Astrof\'{\i}sica e Ci\^{e}ncias do Espa\c{c}o, Universidade do Porto,  Rua das Estrelas, 4150-762 Porto, Portugal}

\author[0000-0003-3627-2561]{Thomas Kallinger} 
\affiliation{Institute of Astrophysics, University of Vienna, 1180 Vienna, Austria}

\author[0000-0002-9573-2567]{Robert R. King} 
\thanks{Now at Met Office, UK}
\affiliation{School of Physics, University of Exeter, Stocker Road, Exeter EX4 4QL, UK}

\author[0000-0003-3020-4437]{Yaguang Li\CNnames{李亚光}} 
\affiliation{Institute for Astronomy, University of Hawai`i, Honolulu, HI 96822, USA}

\author[0000-0002-5648-3107]{Simon J. Murphy} 
\affiliation{Centre for Astrophysics, University of Southern Queensland, Toowoomba, QLD 4350, Australia}

\author[0000-0001-9234-430X]{Jakob L. R{\o}rsted} 
\affiliation{Stellar Astrophysics Centre, Department of Physics and Astronomy, Aarhus University, 8000 Aarhus C, Denmark}
\affiliation{Aarhus Space Centre (SpaCe), Department of Physics and Astronomy, Aarhus University, Denmark}

\author[0000-0002-4879-3519]{Dennis Stello} 
\affiliation{School of Physics, University of New South Wales, Sydney, NSW 2052, Australia}


\begin{abstract}
We have detected solar-like oscillations in the mid K-dwarf \epsind, making it the coolest dwarf to have measured oscillations. The star is noteworthy for harboring a pair of brown dwarf companions and a Jupiter-type planet. We observed \epsind\ during two radial velocity campaigns, using the high-resolution spectrographs HARPS (2011) and UVES (2021). Weighting the time series, we computed the power spectra and established the detection of solar-like oscillations with a power excess located at $5265 \pm 110 \ \micro\hertz$ -- the highest frequency solar-like oscillations so far measured in any star. 
The measurement of the center of the power excess allows us to compute a stellar mass of $0.782 \pm 0.023 \ M_\odot$ based on scaling relations and a known radius from interferometry. We also determine the amplitude of the peak power and note that there is a slight difference between the two observing campaigns, indicating a varying activity level.
 Overall, this work confirms that low-amplitude solar-like oscillations can be detected in mid-K type stars in radial velocity measurements obtained with high-precision spectrographs. 
 
\end{abstract}


\section{Introduction}
\label{sec:intro}

Asteroseismology, the study of stellar oscillations, has flourished in recent years thanks largely to the success of the space missions CoRoT \citep[\textit{Convection, Rotation and planetary Transits},][]{ref:baglin2006a,ref:baglin2006b,ref:auvergne2009}, \textit{Kepler} \citep{ref:borucki2010,ref:kepler_koch,ref:gilliland2010,ref:kepler_chaplin} and TESS \citep[Transiting Exoplanet Survey Satellite][]{ref:ricker2015}. These satellites delivered high-quality photometric times series that are excellent for asteroseismology. However, for bright Sun-like stars, radial velocity observations with a ground-based telescope can deliver better signal-to-noise, due to the much lower stellar granulation background. This is particularly true for cool main-sequence stars, where the oscillation amplitudes are very low \citep{ref:kjeldsen2008, ref:huber2011}.

In the case of solar-like oscillations, the overall frequency pattern is governed by two so-called global asteroseismic parameters namely the large frequency separation (\dnu) and the frequency of maximum power (\numax). These two parameters describe the overall regularity in the frequency comb and the location of the stellar oscillations, respectively, and they are often used to determine fundamental parameters of main-sequence and red-giant stars \citep[see reviews by][]{ref:chaplin2013, ref:hekker2017, ref:garcia2019, ref:Jackiewicz2021}.

Here, we report the detection of solar-like oscillations in the K5\,V star \epsind\ (HD~209100). 
\epsind\ is orbited every ${\sim} 45 \ \mathrm{yr}$ by a jovian mass exoplanet, \epsind b, which has been detected in both RV and astrometry \citep{ref:feng2019}. \epsind b is the nearest cold Jupiter-type planet to Earth and its infrared emission is expected to be sufficiently high to be measured by JWST \citep{ref:feng2023}.

In addition, \epsind\ is part of a hierarchical triple system with the brown dwarfs $\varepsilon$~Indi~Ba and $\varepsilon$~Indi~Bb, which have been extensively studied \citep[see, for example,][]{ref:scholz2003, ref:smith2003, ref:mccaughrean2004, ref:roellig2004, ref:mainzer2007, ref:reiners2007, ref:kasper2009, ref:king2010, ref:cardoso2010, ref:chen2022}. With the current determined dynamical masses of $66.92 \pm 0.36 \ M_\mathrm{Jup}$ and $53.25 \pm 0.29 \ M_\mathrm{Jup}$ for the brown dwarfs \citep{ref:chen2022}, as well as an accurate distance, these are among the best-characterised brown dwarfs \citep{ref:chen2022}. Thereby, the $\varepsilon$ Indi system is highly valuable to put constraints on formation and evolution in the substellar regime \citep{ref:joergens2014,ref:feng2019}.
However, there is disagreement over the age of the system, with estimates spanning a range 0.39--5\,$\giga\mathrm{yr}$ \citep[][]{ref:cannon1970,ref:lachaume1999,ref:rocha2002,ref:barnes2007, ref:kasper2009,ref:king2010,ref:dieterich2018,ref:feng2019, ref:viswanath2021, ref:pathak2021,ref:chen2022}, leaving the age to be firmly pinned down.

In Sec.~\ref{sec:data} we present our radial velocity observations and the spectroscopic analysis. The analysis of the time series with respect to asteroseismology is the topic of Sec.~\ref{sec:analysis}, while the results are presented and discussed in Sec.~\ref{sec:results}. The conclusion follows in Sec.~\ref{sec:conclusion}.


\section{Observational data}
\label{sec:data}

We observed \epsind\ in two intensive radial velocity (RV) campaigns separated in time by about 10 years. The first campaign was carried out with HARPS in 2011, while a UVES campaign followed in 2021. It should be noted that \epsind\ fell in the TESS field in Sectors~1 and 27 and was observed photometrically in the 2-minute-cadence mode for {$\sim$}27\,d each. It was also observed by TESS with 20-s cadence in Sector 68.  However, the oscillations are not visible in any of these data, which is to be expected given their small amplitudes.
\subsection{HARPS RV data}
\label{subsec:harps}

The data were collected over 12 nights in August 2011 with HARPS on the ESO $3.6 \ \meter$ telescope at La Silla Observatory, Chile. Around 40\% of the time was lost due to bad weather. The median exposure time was 30\,s, sometimes increased to 35\,s or 40\,s in poorer conditions.  The exposure time was kept short due to the high expected frequency of the oscillations. The readout time between each exposure was 32\,s, and hence the median sampling time was 62\,s (which corresponds to a Nyquist frequency of about 8000\,\muHz).

\begin{figure*}
\centering
\includegraphics[width=\hsize]{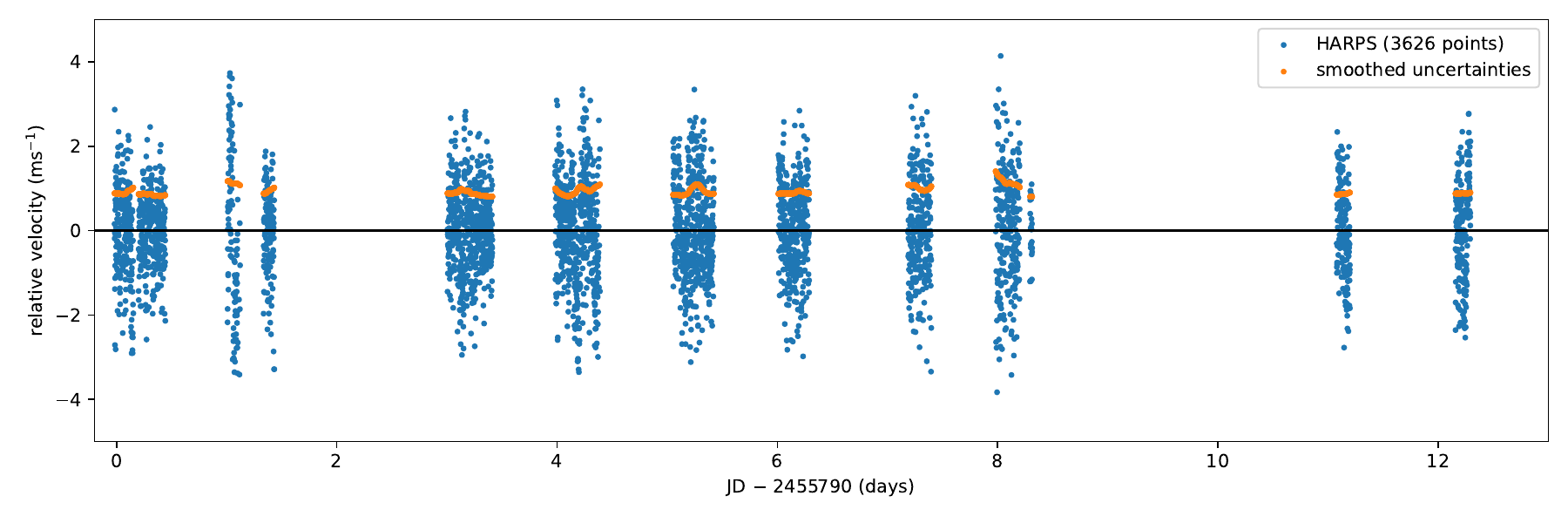}
\includegraphics[width=\hsize]{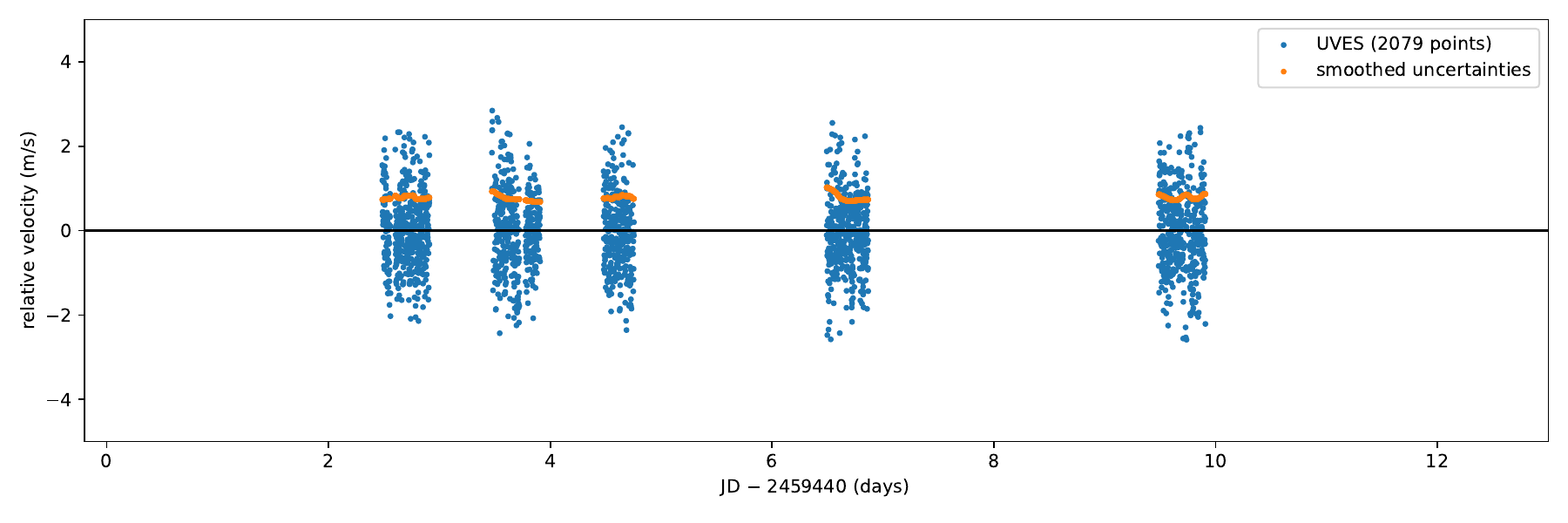}
  \caption{Radial velocity time series of \epsind\ from HARPS (August 2011) and UVES (August 2021). The orange points show the smoothed uncertainties that were measured from the scatter and used for weights when calculating the power spectra (see Sec.~\ref{subsec:ps}). 
          }
     \label{fig:time-series}
\end{figure*}

For this work we used the RVs derived by \citet{ref:trifonov2020} as part of the correction of public HARPS data for systematic errors. A plot of the RV time series can be found in the top panel of Fig.~\ref{fig:time-series}.

\subsection{UVES RV data}
\label{subsec:uves}

In August 2021 we observed \epsind\ during 6 non-consecutive (5+1) nights with UVES on the VLT, using the iodine cell as a wavelength reference. Again, some time was lost due to the weather (around $25\%$), including all of night four. 
The median exposure time was 50\,s, with a readout time between exposures of 21\,s (Nyquist frequency of about 7000\,\muHz).

To obtain the RV time series, all the individual spectra were reduced following the method described in \citet{ref:butler1996,ref:butler2004}. The RV time series is shown in the bottom panel of Fig.~\ref{fig:time-series}.

\subsection{Spectroscopic analysis}
\label{subsubsec:specana}
Some spectra taken as part of the UVES campaign were also used for a spectroscopic analysis. Based on the high signal-to-noise iodine-free template spectrum of \epsind, we were able to derive spectroscopic values of the effective temperature,  surface gravity, [Fe/H], and the alpha enhancement using a classical equivalent-width spectral analysis with the line list from \citet{ref:Slumstrup2019}. 
The results are presented in Table~\ref{tab:spec_params}, with uncertainties internal to the analysis only. 
The auxiliary program \textsc{abundance} within the \textsc{spectrum} software \citep{ref:Gray1994} was used to determine the atmospheric parameters under the assumption of LTE, using ATLAS9 models \citep{ref:Castelli2004} and solar abundances from \citet{ref:Grevesse1998}. We used astrophysical oscillator strengths, with the $\log gf$ values for each absorption line adjusted to yield the solar elemental abundances of \citet{ref:Grevesse1998}. The [$\alpha$/Fe] in this paper is defined as $\frac{1}{4} \cdot \left( \text{ [Ca/Fe] + [Si/Fe] + [Mg/Fe] + [Ti/Fe] } \right)$.

\begin{table}
 \caption{\label{tab:spec_params}Atmospheric parameters derived from the spectroscopic analysis. Note that the quoted uncertainties are internal only (for the abundances; the spread $\pm$ the sensitivity to the other parameters), and $g$ is measured in $\centi\meter\per\second^2$.}
\begin{tabular}{lc}
 \hline \hline
  $T_\mathrm{eff}$      & $4700 \pm 65 \, \kelvin$ \\  
  $\log g$              & $4.50 \pm 0.07$ \\
  $[\mathrm{Fe}/\mathrm{H}]$    & $-0.17 \pm 0.01 \pm 0.03$ \\
  $[\alpha/\mathrm{Fe}]$ & $-0.06 \pm 0.03 \pm 0.05$ \\
  \hline
 \end{tabular}
\end{table}

\epsind\ has been the subject of several previous spectroscopic analyses, dating as far back as 1980 (for an overview, visit for instance \href{http://simbad.cds.unistra.fr/simbad/sim-id?Ident=%403387848&Name=*+eps+Ind&submit=display+all+measurements#lab_meas}{Simbad}). The star is reported to be metal-poor, with a median $\mathrm{[Fe/H]} = -0.16$ dex, which is in excellent agreement with our determination from the UVES spectrum. Median values of the effective temperature ($T_\mathrm{eff} = 4649 \, \kelvin$) and surface gravity ($\log g = 4.54$) are also in agreement with our determined values. Based on abundances from the literature \citep{ref:kollatschny1980, ref:adibekyan2012, ref:delgado2017, ref:luck2018, ref:soto2018, ref:hojjatpanah2019}, it is possible to compute values of the alpha enhancement that can be compared to our value. The median value computed from the literature is $[\alpha/\mathrm{Fe}] = 0.12$ which, in combination with our value, indicates little to no alpha enhancement.


\section{Seismic data analysis}
\label{sec:analysis}

\subsection{Calculation of the power spectrum}
\label{subsec:ps}

\begin{figure}
\centering
\includegraphics[width=\hsize]{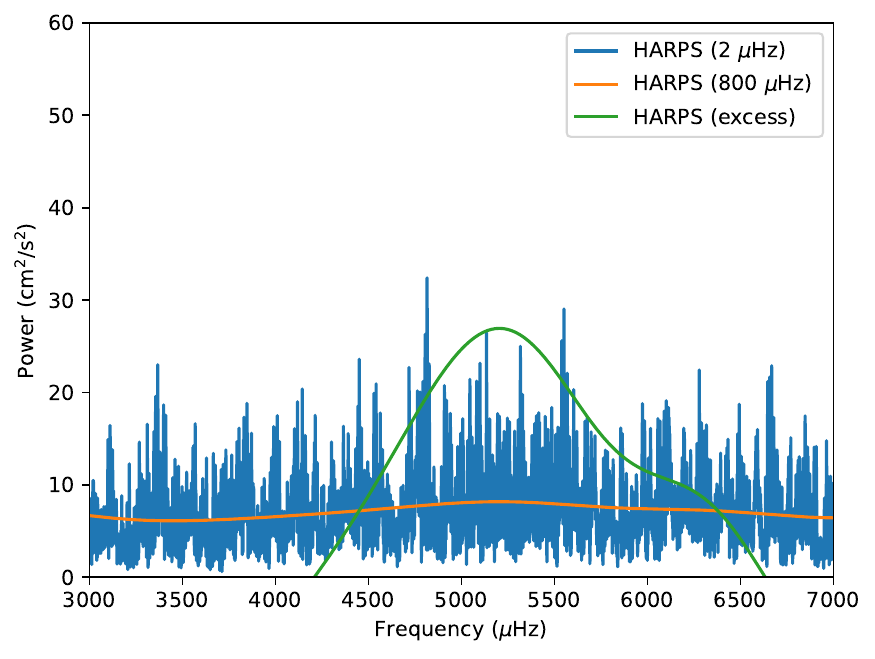}
\includegraphics[width=\hsize]{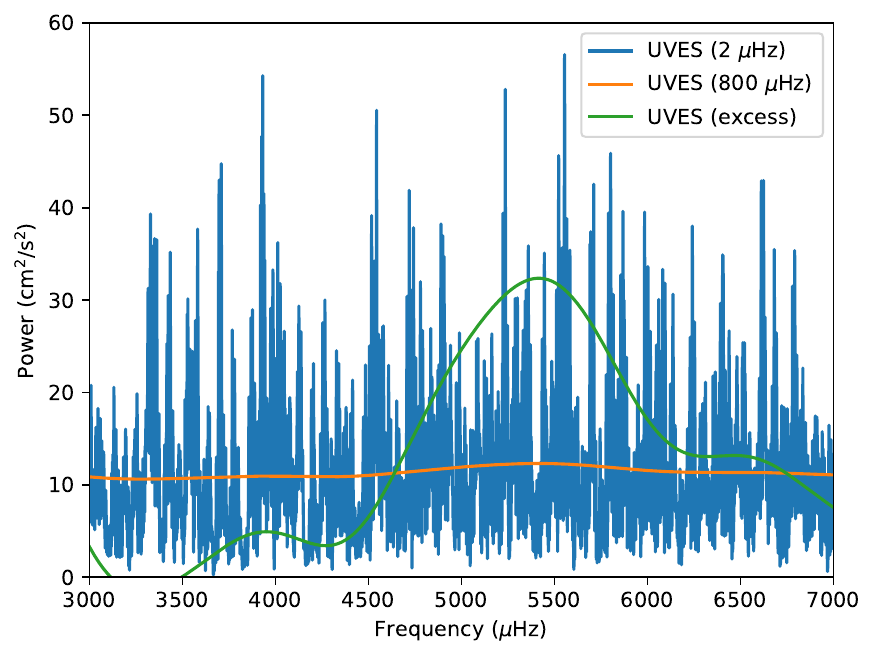}
\includegraphics[width=\hsize]{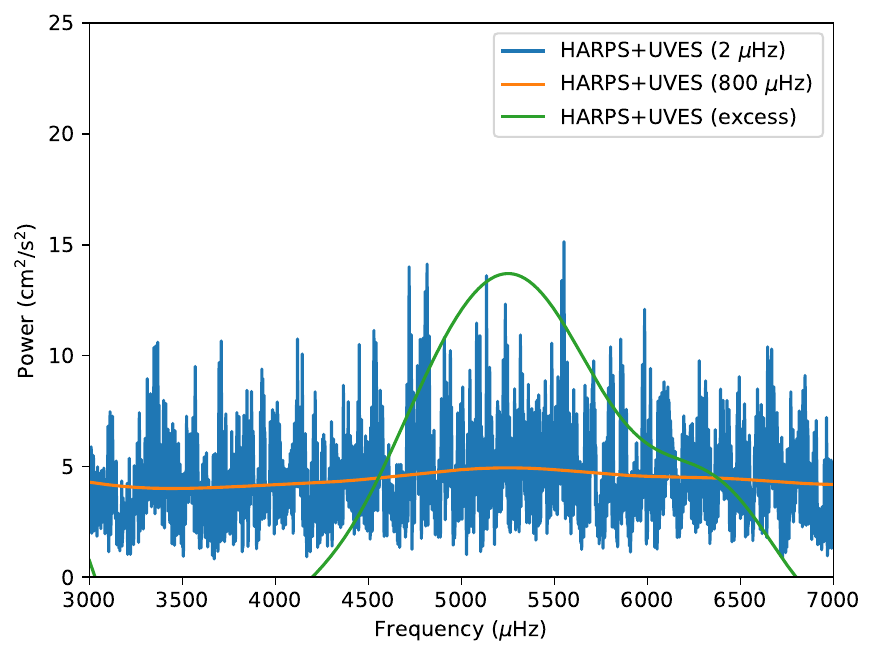}
  \caption{Power spectra of \epsind{} from HARPS (top), UVES (middle), and the combined data (bottom; note the different vertical scale).  In each panel, the blue curve is the power spectrum smoothed over a width of 2\,\muHz, the orange curve is smoothed by a width of 800\,\muHz, and the green curve shows the power excess after subtracting the noise (expanded vertically by a factor of 20 for visibility).
          }
     \label{fig:power-spectra}
\end{figure}

We calculated weighted power spectra using a standard sine-wave fitting technique \citep[see, e.g.][]{ref:kjeldsen1992,ref:frandsen1995,ref:handberg2014}. The weights were taken as the inverse of the measurement uncertainties, which we estimated as the smoothed scatter in the time series (orange points in Fig.~\ref{fig:time-series}). 
The median scatter was $76\,\centi\meter\per\second$ for UVES and $78\,\centi\meter\per\second$ for HARPS.
The slightly smoothed weighted power spectra for the HARPS and UVES observing runs are shown in the top two panels of Fig.~\ref{fig:power-spectra} (blue lines).  

To calculate the power spectrum of the combined HARPS and UVES data we merged them in the time domain, rather than averaging the power spectra, because this allowed us to estimate the spectral window of the combined data \citep[see also][]{ref:bedding2022}.  This combined weighted power spectrum, shown in the bottom panel of Fig.~\ref{fig:power-spectra}, comprises around 70\% power from HARPS and 30\% power from UVES.

\subsection{Location of the power envelope and mode amplitude}

To determine the frequency of maximum oscillation power, \numax, and to estimate the amplitude of the oscillations, we followed the method described by \citet{ref:kjeldsen2005, ref:kjeldsen2008}. This involves the following steps: (i)~remove the structure from individual p modes in the power spectrum by heavily smoothing with a Gaussian with a full-width at half maximum of 800\,\muHz{} ($\sim4\Dnu$, with \Dnu\ being the expected large frequency separation; orange curves in Fig.~\ref{fig:power-spectra}); (ii)~convert to power density by multiplying by the effective length of the observing run (calculated as the reciprocal of the area under the spectral window in power); and (iii)~fit and subtract the background noise using a linear fit to the area surrounding the power envelope (orange curves in Fig.~\ref{fig:power-spectra}). From this, \numax\ can be found as the peak position.

A last step is needed to extract the mode amplitude, namely (iv)~multiply by 4.09 (the effective number of modes per order) and take the square root, in order to convert to amplitude per oscillation mode.


\section{Results}
\label{sec:results}

\subsection{Measurement of \numax}
\label{subsec:numax_measure}

The determined center of the power excess (\numax) as well as the amplitude of the central radial mode (having degree $\ell = 0$) can be found for the HARPS, UVES and combined power spectra in Table~\ref{tab:seis_params}. The main source of uncertainty for \numax\ and the peak amplitude is the noise background. We estimated the error bars for the three excess power estimates by making simple simulations with a known input, using the same spectral window and signal-to-noise ratio as the real data. 
In the following, we chose to use the \numax\ from the combined power spectrum as this is based on the most data, which is a choice in line with our previous multi-site campaigns to detect solar-like oscillations \citep[e.g.][]{ref:kjeldsen2005, ref:bedding2006_nuInd, ref:bedding2007_betaHyi_dnu, ref:arentoft2008_procyon}.

As an effect of the non-zero exposure time, the mode amplitudes will decrease slightly and the frequency of maximum power will shift to lower frequencies; an effect sometimes referred to as apodization \citep[for the effect on mode amplitudes see, e.g.,][]{ref:chaplin2011ApJ,ref:kallinger2014}. To understand this effect, all three data sets (HARPS, UVES and the combined) were simulated 100 times with the actual exposure times and using the individual statistical weights. The effect for the three data sets---based on simulating a p-mode excess power envelope at 5250 microHz---is that, to account for the non-zero exposure times, the measured peak amplitudes should be increased by $4.2\%$ (HARPS), $12.2\%$ (UVES) and 6.8\% (combined), while the \numax\ values should be increased by 0.1\% (HARPS), 0.4\% (UVES) and 0.2\% (combined). While the shift of \numax\ is well below the uncertainty, the effect on the amplitude is relatively larger, although still below the uncertainty of the amplitude estimate. These corrections have been applied in Table~\ref{tab:seis_params}.

\begin{table*}
\caption{Frequency of maximum oscillation power (\numax) and amplitudes of the central $\ell = 0$ mode of \epsind\ (corrected for apodization).}
\label{tab:seis_params}
\centering
\begin{tabular}{l c c c}
\hline\hline                 
Quantity & Symbol & Dataset & Value \\    
\hline                        
   Frequency of maximum power & $\nu_\mathrm{max}$ & UVES      & $5440 \pm 170  \, \micro\hertz$ \\
                              &                    & HARPS     & $5210 \pm 145 \, \micro\hertz$ \\ 
                              &                    & Combined  & $5265 \pm 110 \, \micro\hertz$ \\ 
   Peak amplitude & $A (\ell = 0 \mathrm{~at~peak})$     & UVES      & $3.1 \pm 0.9 \, \centi\meter\per\second$ \\ 
                  &                                     & HARPS     & $3.6 \pm 0.7 \, \centi\meter\per\second$ \\
                  &                                     & Combined  & $3.4 \pm 0.6 \, \centi\meter\per\second$ \\
\hline                                   
\end{tabular}
\end{table*}

\subsection{Mass of \epsind}

As mentioned in the Introduction, it is common to estimate the mass of a star with solar-like oscillations from \numax\ and \Dnu\ using the asteroseismic scaling relations. However, if a reliable luminosity or radius is available, the mass can be estimated from \numax\ without requiring \Dnu, as has been done in studies of red giants \citep{ref:stello2008, ref:hon2021_QLP}.
Based on the frequency of maximum power (Table~\ref{tab:seis_params}), the effective temperature (Table~\ref{tab:spec_params}) as well as the computed radius from combining the distance with the measured angular diameter, we can calculate the mass of \epsind\ using the scaling relations \citep[see e.g.][]{ref:sca_brown,ref:sca_kjeldsen}. Specifically, the mass can be found as  
\citep{ref:stello2008, ref:kallinger2010}:
\begin{equation}
    \label{eq:mass_numax}
    \frac{M}{M_\odot} \approx \left( \frac{\nu_\mathrm{max}}{\nu_{\mathrm{max},\odot}} \right)
                              \left( \frac{R}{R_\odot} \right) ^2
                              \left( \frac{T_\mathrm{eff}}{T_{\mathrm{eff},\odot}} \right) ^{0.5} \ ,
\end{equation}
with suitable solar reference values. In this work we adopt $\nu_{\mathrm{max},\odot} = 3090 \pm 30 \ \micro\hertz$ \citep{ref:huber2011} and $T_{\mathrm{eff},\odot} = 5772.0 \pm 0.8 \ \kelvin$ \citep{ref:prsa2016}. To obtain the radius of \epsind\ we used the distance from the astrophysical parameters table of Gaia DR3 \citep{ref:creevey2023}: $d = 3.648 ^{+0.022}_{-0.009}$ pc (\textsc{distance\_gspphot}). Combining this with the limb-darkened angular diameter of $\theta_\mathrm{LD} = 1.817 \pm 0.013$ mas \citep{ref:rains2020} gives a radius of $0.713 \pm 0.006 \ R_\odot$, in excellent agreement with the value found by \citet{ref:rains2020}. Using Eq.~\ref{eq:mass_numax} yields a stellar mass of $0.782 \pm 0.023 \ M_\odot$.
This is in agreement with the mass of $0.762 \pm 0.038$ found by \citet{ref:demory2009} using the interferometric radius and mass--luminosity relationship from \citet{ref:xia2008}.


\subsection{Mode amplitude variation with the activity cycle}
\label{sec:activity}

We see indications of a difference between the mode amplitudes observed in the HARPS and UVES time series (see Table~\ref{tab:seis_params}). This is not surprising as the two time series are separated by a time period of a decade, and the mode amplitudes in the Sun are known to vary over its 11-yr activity cycle \citep{ref:chaplin2000, ref:kjeldsen2008, ref:kim2022}. 

\begin{figure}
\centering
\includegraphics[width=\hsize]{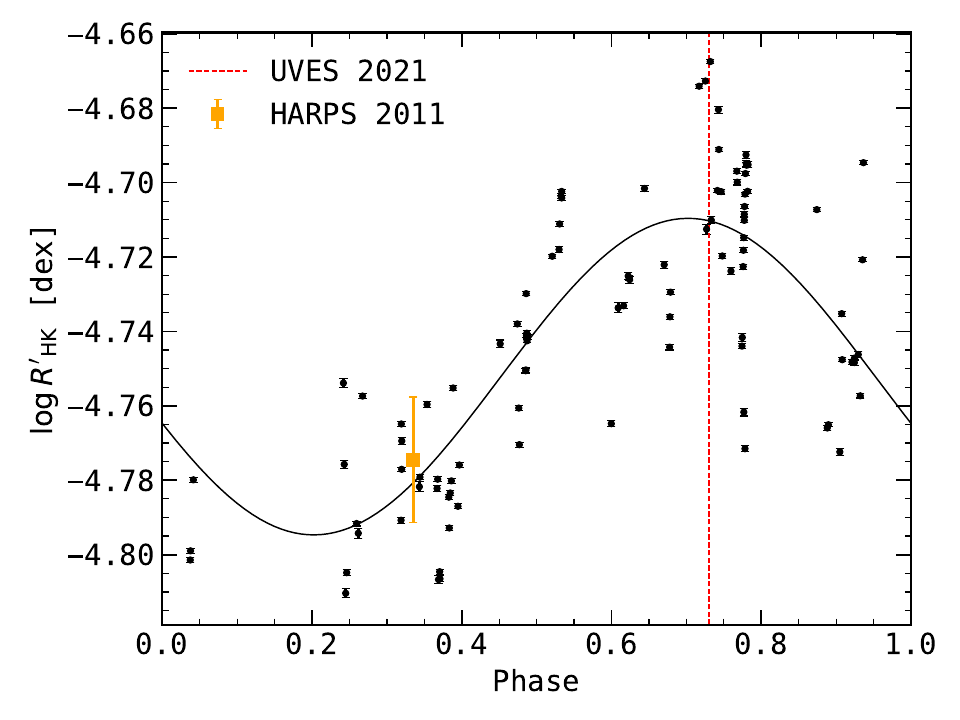}
  \caption{Phase-folded activity cycle of \epsind\ from $\log R^{\prime}_{\rm HK}$. The black dots show the HARPS observations with our seismic campaign marked in yellow (error bar given by the standard deviation of the observations taken during the campaign). The red dotted line indicates the time of our UVES seismic observations, where the Ca-II H\&K lines were not measured, and the solid black line represent a simple sinusoidal variation of $\log R^{\prime}_{\rm HK}$ with a period of $2600$ days.}
     \label{fig:activity}
\end{figure}

In fact, making use of 4293 HARPS archival data obtained for \epsind\ between 2003 and 2016, binned to 112 nights, we can place our observations in the context of the activity cycle of \epsind. From the HARPS archival data, we obtained the $\log R^{\prime}_{\rm HK}$ values \citep[following][]{ref:gomes2018,ref:gomes2021} and used those to model the activity cycle of \epsind. Figure~\ref{fig:activity} shows the variation of $\log R^{\prime}_{\rm HK}$ phase-folded onto the estimated cycle period of $\sim\!2600$ days. The scatter of the HARPS observations about the sinusoidal model is caused by rotational activity variations.

As can be seen from the figure, we happened to observe \epsind\ both near the minimum (HARPS) and near the maximum (UVES) of activity. The difference between our two measurements of the oscillation amplitude is only marginally significant (Table~\ref{tab:seis_params}) but it is noteworthy that it is consistent with expectations of an inverse correlation with activity.

\section{Conclusions}
\label{sec:conclusion}

We have detected solar-like oscillations in the K5 dwarf \epsind. 
Our measured value of \numax\ of $5265 \pm 110 \, \micro\hertz$ is the highest so far found for solar-like oscillations, surpassing Kepler-444 (K0~V; 4540\,\muHz; \citealt{ref:campante2015}), $\tau$~Ceti (G8~V; 4490\,\muHz; \citealt{ref:teixeira2009}) and  $\alpha$~Cen~B (K1~V; 4090\,\muHz; \citealt{ref:kjeldsen2005}).
Combining this detection with a known distance and angular diameter, we estimate the mass of \epsind\ to be $0.782 \pm 0.023 \ M_\odot$, in agreement with literature values.
In addition, we see indications of the expected inverse correlation between the activity cycle and the observed mode amplitudes.
This result paves the way for detections of solar-like oscillations in more early-to-mid K-dwarfs.


\acknowledgments
      Based on observations collected at the European Southern Observatory under ESO programmes 087.D-0511 and 105.20CY. 
      This work was supported by a research grant (42101) from VILLUM FONDEN as well as The Independent Research Fund Denmark's Inge Lehmann  program (grant  agreement  no.:  1131-00014B). Funding for the Stellar Astrophysics Centre was provided by The Danish National Research Foundation (grant agreement no.: DNRF106).
      TRB is supported by an Australian Research Council Laureate Fellowship (FL220100117).
      TLC is supported by Funda\c c\~ao para a Ci\^encia e a Tecnologia (FCT) in the form of a work contract (CEECIND/00476/2018).
      J.G.S. acknowledges support from the Funda\c c\~ao para a Ci\^encia e a Tecnologia (FCT) the research grants UIDB/04434/2020 \& UIDP/04434/2020 and 2022.04416.PTDC. This work was co-funded by the European Union (ERC, FIERCE, 101052347). Views and opinions expressed are however those of the author(s) only and do not necessarily reflect those of the European Union or the European Research Council. Neither the European Union nor the granting authority can be held responsible for them.
      CA received funding from the European Research Council (ERC) under the Horizon Europe programme (Synergy Grant agreement No.\ 101071505: 4D-STAR).
      DS is supported by the Australian Research Council Discovery Grant (DP190100666).
      This research has made use of the SIMBAD database, operated at CDS, Strasbourg, France.
      This work has made use of data from the European Space Agency (ESA) mission Gaia (\url{https://www.cosmos.esa.int/gaia}), processed by the Gaia Data Processing and Analysis Consortium (DPAC, \url{https://www.cosmos.esa.int/web/gaia/dpac/consortium}). Funding for the DPAC has been provided by national institutions, in particular the institutions participating in the Gaia Multilateral Agreement. We thank the referee for helpful comments.

\facilities{ESO:3.6m (HARPS), VLT:Kueyen (UVES)} 

\clearpage

 \newcommand{\noop}[1]{}


\end{document}